\documentclass[preprint,aps,prl,nofootinbib,superscriptaddress]{revtex4-1}
\usepackage[utf8]{inputenc}

\usepackage{amsmath}    
\usepackage{graphicx}   
\usepackage{verbatim}   
\usepackage{color}      
\usepackage{epstopdf}  
\usepackage{upgreek} 
\usepackage[hidelinks]{hyperref}   
\usepackage{textcomp} 
\usepackage[separate-uncertainty=true, multi-part-units=single]{siunitx}
\usepackage{dcolumn}
\usepackage[capitalise]{cleveref}

\newcommand{\MuMax}{MuMax$^{3}$}
\newcommand{\Ne}{N\'{e}el}

\begin{document}

\title{Stabilizing chiral spin-structures via an alternating Dzyaloshinskii–Moriya interaction}
\author{Juriaan Lucassen}
\email{j.lucassen@tue.nl}
\affiliation{Department of Applied Physics, Eindhoven University of Technology, P.O. Box 513, 5600 MB Eindhoven, the Netherlands}
\author{Mari\"{e}lle J. Meijer}
\affiliation{Department of Applied Physics, Eindhoven University of Technology, P.O. Box 513, 5600 MB Eindhoven, the Netherlands}
\author{Mark C.H. de Jong}
\affiliation{Department of Applied Physics, Eindhoven University of Technology, P.O. Box 513, 5600 MB Eindhoven, the Netherlands}
\author{Rembert A. Duine}
\affiliation{Department of Applied Physics, Eindhoven University of Technology, P.O. Box 513, 5600 MB Eindhoven, the Netherlands}
\affiliation{Institute for Theoretical Physics, Utrecht University, Princetonplein 5, 3584 CC Utrecht, the Netherlands}
\author{Henk J.M. Swagten}
\affiliation{Department of Applied Physics, Eindhoven University of Technology, P.O. Box 513, 5600 MB Eindhoven, the Netherlands}
\author{Bert Koopmans}
\affiliation{Department of Applied Physics, Eindhoven University of Technology, P.O. Box 513, 5600 MB Eindhoven, the Netherlands}
\author{Reinoud Lavrijsen}
\email{r.lavrijsen@tue.nl}
\affiliation{Department of Applied Physics, Eindhoven University of Technology, P.O. Box 513, 5600 MB Eindhoven, the Netherlands}

\date{\today}

\begin{abstract}
The stabilization of chiral magnetic spin-structures in thin films is often attributed to the interfacial Dzyaloshinskii–Moriya interaction (DMI). Very recently, however, it has been reported that the chirality induced by the DMI can be affected by dipolar interactions. These dipolar fields tend to form \Ne{} caps, which entails the formation of a clockwise chirality at the top of the film and a counterclockwise chirality at the bottom. Here, we show that engineering an alternating DMI that changes sign across the film thickness, together with the tendency to form Neel caps, leads to an enhanced stability of chiral spin-structures. Micromagnetic simulations for skyrmions demonstrate that this can increase the effective DMI in a prototypical [Pt/Co/Ir] multilayer system by at least \SI{0.6}{mJ.m^{-2}}. These gains are comparable to what has been achieved using additive DMI, but more flexible as we are not limited to a select set of material combinations. We also present experimental results: by measuring equilibrium domain widths we quantify the effective DMI in [Pt/Co/Ir] multilayer systems typically used for skyrmion stabilization. Upon introducing an alternating DMI we demonstrate changes in the effective DMI that agree with our simulations. Our results provide a route towards enhancing the stability of chiral spin-structures that does not rely on enlarging the chiral interactions.
\end{abstract}
\maketitle
Magnetic skyrmions are whirling chiral spin-structures that can be as small as a few~\si{nm}.~\cite{Nagaosa2013,Wiesendanger2016,Fert2017,doi:10.1063/1.5048972} Because of their topological protection, they are extremely stable magnetic quasiparticles that might find their use in many applications such as magnetic racetrack memory.~\cite{Nagaosa2013,PhysRevB.93.214412,Fert2017,Wiesendanger2016,doi:10.1063/1.5048972,Fert2013} Skyrmions are typically stabilized by the Dzyaloshinskii–Moriya interaction (DMI), which originates from a global inversion symmetry breaking in combination with spin-orbit coupling.~\cite{PhysRev.120.91,DZYALOSHINSKY1958241} Although skyrmions exist in many systems,~\cite{Muhlbauer915,Yu2010,Heinze2011,Wiesendanger2016,Woo2016,Luchaire_skyrmion,Boulle2016} there is a great interest in skyrmions stabilized in ultra-thin ferromagnets. Their stabilization is achieved through the interfacial DMI from a symmetry-breaking interface between an ultra-thin ferromagnet and a heavy metal.~\cite{Bogdanov2001,fert1990,CREPIEUX1998341} In these ultrathin systems, the magnetic properties can be tailored for specific applications by varying the magnetic layer thicknesses and interfaces.

Unfortunately, the DMI is often not large enough to stabilize magnetic skyrmions at room temperature. To compensate for this, the magnetic volume is usually increased to enhance the thermal stability and reduce the skyrmion energy.~\cite{Woo2016,Luchaire_skyrmion,Boulle2016,Soumyanarayanan2017} The concomitant increase of dipolar interactions, however, has been shown both theoretically and experimentally to compete with the DMI leading to a non-uniform magnetic chirality across the thickness of the layers.~\cite{Hubert1998,Lucassen2019,doi:10.1002/adma.201807683,PhysRevLett.122.237201,Montoya2017,Legrand2018,Lemesh2018,Legrand2018a,Dovzhenko2018,Fallon2019} This is considered detrimental for applications because most of the functionality relies on the uniform chirality of a skyrmion across the thickness of the multilayer system.\footnote{As the chirality of the skyrmions varies between individual repeats of the multilayer system these are not skyrmions in the strictest sense.~\cite{Nagaosa2013,Wiesendanger2016,Fert2017,doi:10.1063/1.5048972}}  

\begin{figure}[!t]
	\includegraphics[width=\columnwidth]{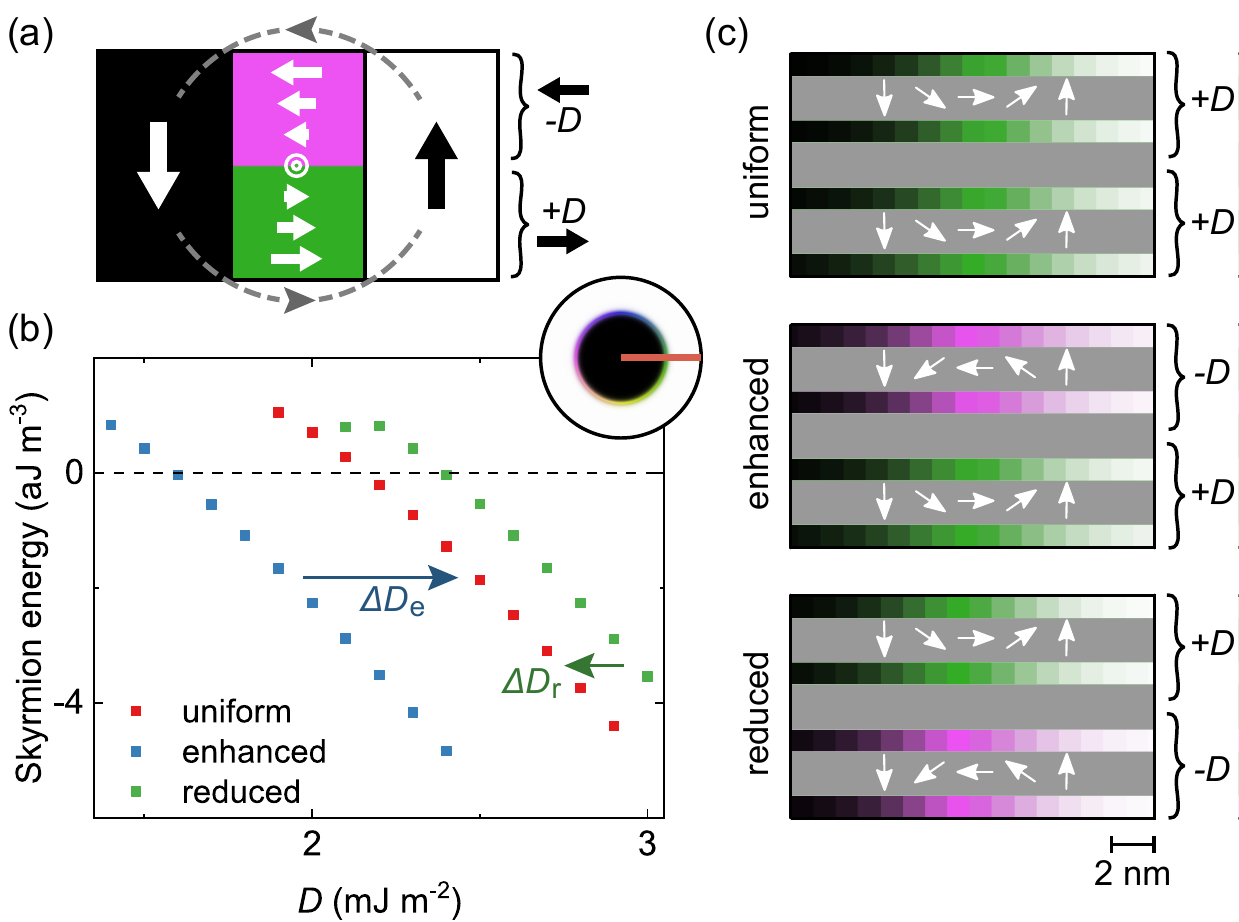}
	\caption{\label{fig:out2_over}(a) Schematic overview of the physical interactions exploited to increase the stability of chiral spin-structures. The dipolar fields (grey) create two \Ne{} caps in the domain wall. Halfway through the layer, the sign of the DMI $D$ is reversed such that the effective fields of the dipolar interactions are everywhere aligned with the effective field of the DMI (black). (b) Skyrmion energy relative to the uniformly magnetized state as a function of $D$ for the three DMI configurations shown in (c). The arrows indicate the effective DMI gain ($\Delta D_\mathrm{e}$) and loss ($\Delta D_\mathrm{r}$). The only time a datapoint is included is when both the uniformly magnetized and skyrmion state are (meta)stable. Inset: simulation geometry with a skyrmion in a confined dot, where the red horizontal line indicates the profile of a domain wall such as shown in (a) and (c). (c) Stack configurations for the three different DMI configurations with $N=4$, where we show the resulting cross-section of a skyrmion profile using white arrows for the bottom and top layer for $N=4$ and $D=2.2$~\si{mJ.m^{-2}}.}
\end{figure}

On the other hand, despite their negative effect on the magnetic chirality, even without the DMI dipolar interactions are able to stabilize so-called dipolar skyrmions.~\cite{Hrabec2017,Bellec_2010,2019arXiv190908909M} This occurs through the formation of \Ne{} caps, which is the formation of a clockwise chirality at the top of the film and a counterclockwise chirality at the bottom. Inspired by this, we suggest here to combine the formation of \Ne{} caps with a layer-dependent alternating DMI to enhance the stability of chiral spin-structures as is shown schematically in~\cref{fig:out2_over}a for a magnetic domain wall. The dipolar fields introduce \Ne{} caps with a clockwise (CW) \Ne{} wall at the top of the film, and a counterclockwise (CCW) \Ne{} wall at the bottom. For a uniform DMI this leads to a competition with the DMI across half the stack. Therefore, we intentionally reverse the sign of the DMI halfway through the system, such that in both halves of the stack the DMI field points in the same direction as the dipolar fields, which leads to a reduction in both the domain wall and skyrmion energy. In the first part of this Letter, we investigate this principle using \MuMax{} based micromagnetic simulations~\cite{Vansteenkiste2014} and demonstrate that it leads to significant increases in the effective DMI by comparing skyrmion energies for different DMI configurations. Thereafter, we also present experimental results on the effect of modifying the DMI in a multilayer [Pt/Co/Ir] system. Upon changing the DMI configuration we find almost a factor $2$ increase in the effective DMI after accounting for growth-induced variations in the magnetic parameters through an averaging approach, which we verify with micromagnetic simulations. This proves that \Ne{} caps can be exploited to significantly increase the stability of chiral spin-structures and opens up a way to tailor them by modifying magnetic interactions on a layer-by-layer basis.

We investigate the behaviour of confined magnetic skyrmions in a \SI{256}{nm} diameter circular dot (inset \cref{fig:out2_over}b) with \MuMax{} using [NM($2$)/FM($1$)]x$N$ systems, with $N$ repeats of a~\SI{1}{nm} thick ferromagnetic layer (FM) sandwiched in between~\SI{2}{nm} thick non-magnetic (NM) spacer layers. The magnetic parameters used for these calculations correspond to the experimental parameters of the prototypical [Pt/Co/Ir] systems experimentally investigated later in this Letter. 
More details on the simulations can be found in supplementary note I.
In~\cref{fig:out2_over}b the energy of the skyrmion state with respect to the uniformly magnetized state is plotted as a function of the DMI $D$ for $3$ different DMI configurations which are indicated in~\cref{fig:out2_over}c: i) a uniform configuration, where the DMI is equal across all layers, ii) an enhanced DMI configuration, where the sign of the DMI aligns along the internal dipolar fields everywhere in the stack, and iii) a reduced DMI configuration where the sign of the DMI is always aligned anti-parallel to the dipolar fields. The enhanced and reduced DMI configurations lead to the formation of a thickness dependent chirality by the introduction of \Ne{} caps. For all DMI configurations, the skyrmion energy decreases with increasing DMI as expected. In addition, the enhanced DMI configuration leads to a significantly reduced skyrmion energy, and the reduced configuration to an increase in the skyrmion energy. This is completely in line with the simple picture sketched in~\cref{fig:out2_over}a. Specifically, as indicated in~\cref{fig:out2_over}b with the DMI gain $\Delta D_\mathrm{e}$, the $D$ required to obtain a skyrmion whose energy is lower than the uniformly magnetized state decreases by \SI{0.6}{mJ.m^{-2}} upon introducing the enhanced configuration. This is a massive increase in stability and comparable in magnitude to the gains obtained when utilizing an effect like additive DMI.~\cite{Luchaire_skyrmion,Soumyanarayanan2017,Yang2018} Moreover, as we are not bound to the small set of material systems with a large additive DMI this should be more widely applicable. In supplementary note II we additionally show that while $\Delta D_\mathrm{e}$ does vary with $N$, the presented behaviour remains qualitatively identical.

\begin{figure}
	\includegraphics[width=\columnwidth]{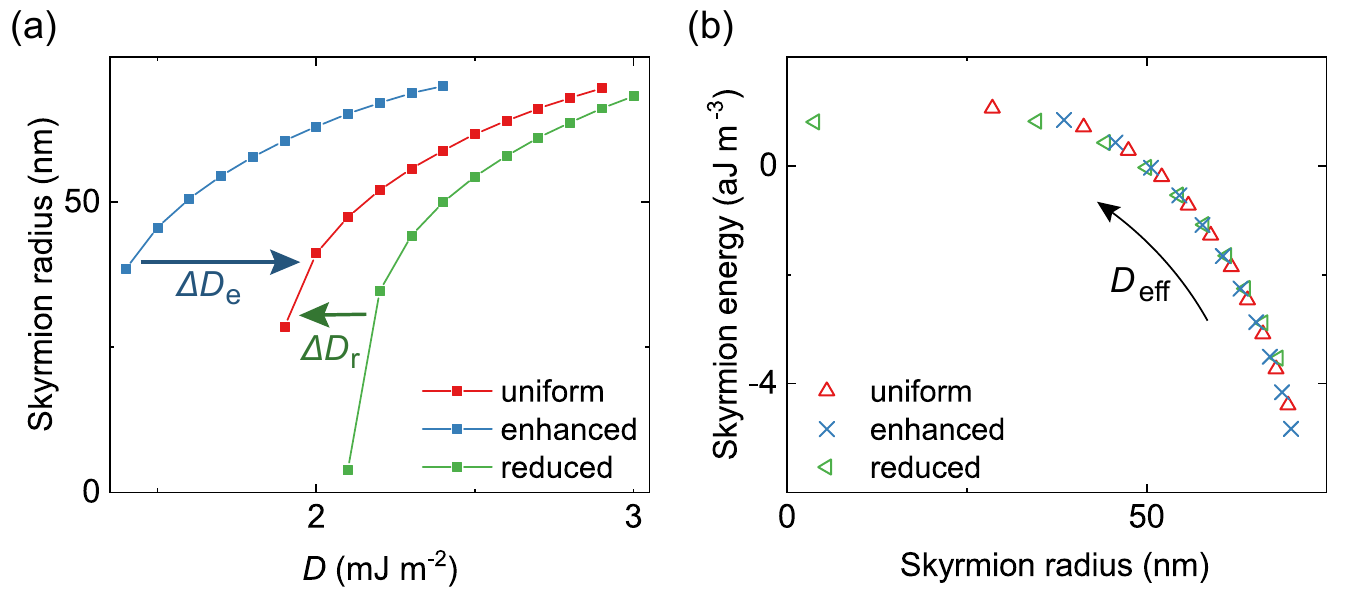}
	\caption{\label{fig:out2_size}(a) Skyrmion radii as function of $D$ for the three different DMI configurations with $N=4$. The radius is determined from the position where the magnetization along the out-of-plane axis changes sign. (b) Skyrmion energy as a function of skyrmion radius for the three different DMI configurations with $N=4$. The arrow indicates the direction of increasing $D_\mathrm{eff}$.}
\end{figure}

Introducing these different DMI configurations also has a profound effect on the skyrmion radius as shown in~\cref{fig:out2_size}a.\footnote{In supplementary note III we also show the results for the simulations under an applied magnetic field, where the enhanced DMI configuration leads to a larger field stability.} In agreement with the results found in literature, an increase in $D$ leads to an increase in the skyrmion radius.~\cite{PhysRevB.88.184422,Luchaire_skyrmion} We can understand this by thinking of a skyrmion as an out-of-plane (OOP) magnetized core enclosed by a domain wall.~\cite{PhysRevLett.114.177203,PhysRevB.88.184422} As $D$ increases the domain wall energy decreases, resulting in a skyrmion that can expand to enhance the dipolar coupling of the core to the annulus. The same mechanism explains the behaviour for the three different DMI configurations; the skyrmion becomes bigger (smaller) when introducing the enhanced (reduced) DMI configuration because the domain wall energy decreases (increases). In supplementary note IV we further illustrate that the confinement effect of the simulated dot does not qualitatively affect the presented behaviour.


We will now try to understand the effect of the different DMI configurations in a more general way. Combined, the behaviour depicted in~\cref{fig:out2_over}b and~\cref{fig:out2_size}a suggests some form of universality. All the curves show qualitatively the same behaviour as a function of $D$, apart from the shifted $D$ values indicated by the arrows of DMI gain ($\Delta D_\mathrm{e}$) and loss ($\Delta D_\mathrm{r}$). This can be understood by considering the effect of the dipolar fields: as suggested by Lemesh \textit{et al.},~\cite{Lemesh2018} the dipolar fields can be included as an effective DMI because both components introduce an effective in-plane magnetic field in the domain wall. 
We can thus introduce an effective DMI $D_\mathrm{eff}=D+\Delta D_\mathrm{e,r}$ (see~\cref{fig:out2_over}b) for the enhanced and reduced configurations. In this specific case, the enhanced configuration behaves as a system with an effective DMI $D_\mathrm{eff}$ that is $\Delta D_\mathrm{e} \approx + 0.6$~\si{mJ.m^{-2}} larger than the DMI $D$ because of the additive effects of the DMI and dipolar interactions. Conversely, the reduced configuration has a smaller effective DMI with $\Delta D_\mathrm{r} \approx - 0.2$~\si{mJ.m^{-2}}. A consequence of this universality is shown in~\cref{fig:out2_size}b, where the skyrmion energy is plotted as a function of skyrmion radius for the three different DMI configurations. The simulations for the different DMI configurations collapse on the same curve. Thus, although for each configuration the $D$ needed to obtain a particular energy/radius is different, the relationship between the two remains unaffected and can be described by an $D_\mathrm{eff}$ that has a DMI-configuration dependent contribution.
Last, in supplementary note V we demonstrate that the introduction of \Ne{} caps can in some cases lead to non-circular skyrmions to accommodate both the DMI and dipolar interactions.

\begin{figure}[!t]	
	\includegraphics[width=\columnwidth]{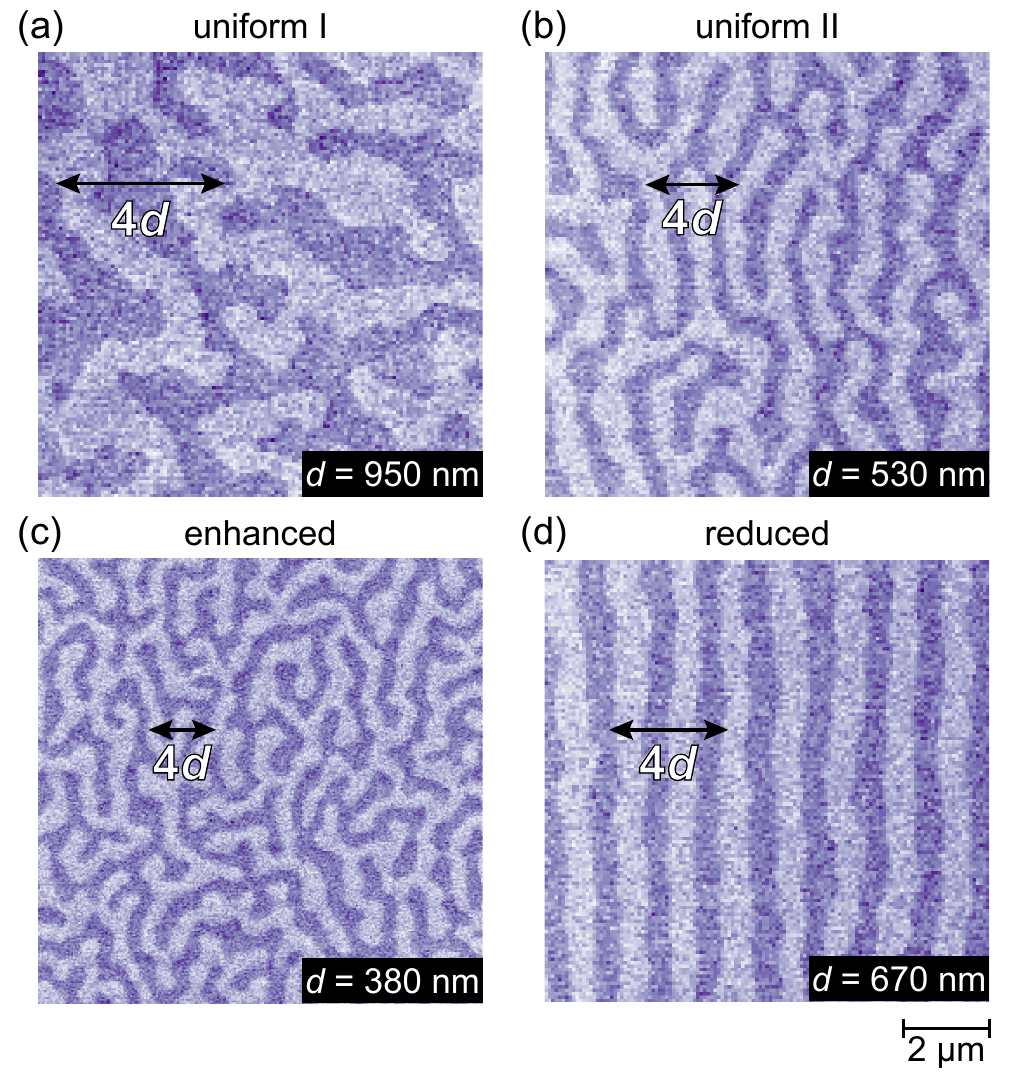}
	\caption{\label{fig:out2_MFM}MFM images of a demagnetized domain state for (a) the uniform I stack, (b) the uniform II stack,  (c) the enhanced stack, and (d) the reduced stack. The scale bar at the bottom right holds for all figures. The arrows indicate the average domain width $d$, which we also show in the bottom right of each scan. 
	}
\end{figure}

In the previous paragraphs we have introduced the unique ability of a layer-dependent DMI configuration to enhance the stability of skyrmions. This part presents experimental evidence which shows that $D_\mathrm{eff}$ can be tailored by modifying the DMI on a layer-by-layer basis. To demonstrate this, instead of skyrmions we shift our attention to domain walls for their much easier experimental access, and their fully analogous underlying physics. They allow us to accurately quantify $D_\mathrm{eff}$ because the domain width $d$ in magnetic multilayers is determined by the competition between the domain wall energy and dipolar interactions between the domains.~\cite{Draaisma1987,Suna1986,Lemesh2017,Boulle2016,Luchaire_skyrmion,Woo2016,Lemesh2018,Legrand2018} Here, we use the accurate stripe domain model by Lemesh~\textit{et al.}~\cite{Lemesh2017} to determine $D_\mathrm{eff}$ in four different Ta($4$)/Pt($2$)/X/Ta($4$) systems (thicknesses in parentheses in \si{nm}) to investigate the effect of the different DMI configurations. X for each stack is given by: \vspace{2.5mm} \newline{} 
\textbf{Uniform I:} $\overbrace{\mathrm{[Pt}(1)\mathrm{/Co}(1)\mathrm{/Ir}(1)\mathrm{]}}^{+D}$x$4$, \vspace{2.5mm} \newline{} 
\textbf{Uniform II:} $\overbrace{\mathrm{[Ir}(1)\mathrm{/Co}(1)\mathrm{/Pt}(1)\mathrm{]}}^{-D}$x$4$, \vspace{2.5mm} \newline{} 
  \textbf{Enhanced:} $\overbrace{\mathrm{[Pt}(1)\mathrm{/Co}(1)\mathrm{/Ir}(1)\mathrm{]}}^{+D}$x$2$$\overbrace{\mathrm{[Ir}(1)\mathrm{/Co}(1)\mathrm{/Pt}(1)\mathrm{]}}^{-D}$x$2$, \vspace{2.5mm} \newline{}
  \textbf{Reduced:} $\overbrace{\mathrm{[Ir}(1)\mathrm{/Co}(1)\mathrm{/Pt}(1)\mathrm{]}}^{-D}$x$2$$\overbrace{\mathrm{[Pt}(1)\mathrm{/Co}(1)\mathrm{/Ir}(1)\mathrm{]}}^{+D}$x$2$. \vspace{2.5mm} \newline{} 
Pt and Ir were chosen because of their opposite interfacial DMI signs such that they favour CCW \Ne{} walls for a Pt/Co/Ir stacking ($+D$) and CW \Ne{} walls for an Ir/Co/Pt stacking ($-D$).~\cite{Ma2018,Tacchi2017,doi:10.1021/acs.nanolett.6b01593,Finizio2019,Luchaire_skyrmion} The two different uniform stacks were fabricated to investigate the contributions of the stacking order. In~\cref{fig:out2_MFM} the resulting domain patterns after demagnetization are shown.  These were recorded using magnetic force microscopy (MFM) measurements and the domain width $d$ was extracted through a Fourier analysis. The experimental details can be found in supplementary note I.
\begin{table*}
\caption{\label{tab:out2_DMI_extract}%
Analysis of the domain widths $d$ for the four different stacks extracted from~\cref{fig:out2_MFM}. From these values the effective DMI $|D_\mathrm{eff}|$ was calculated using the accurate stripe domain model with $A=10$~\si{pJ.m^{-1}}.~\cite{Lemesh2017} The uniform averaged stack is a hypothetical stack with the average magnetic parameters of the uniform I and II stacks.}
\begin{tabular}{l cccccc }
\hline\hline
\noalign{\vskip 1mm}  
         &   $d$ (\si{nm}) & $M_\mathrm{s}$ (\si{MA.m^{-1}}) & $K$ (\si{MJ.m^{-3}}) & $|D_\mathrm{eff}|$ (\si{mJ.m^{-2}})  \\   
     \noalign{\vskip 1mm} 
Uniform I & \num{9.5 +- 0.4e2} & \num{0.93 +- 0.03}& \num{1.02 +- 0.04} & \num{1.3 +- 0.1} \\
Uniform II & \num{5.3 +- 0.1e2} & \num{1.30 +- 0.05}& \num{1.60 +- 0.08} & \num{0.6 +- 0.2} \\
Uniform averaged & - & \num{1.12 +- 0.06}& \num{1.31 +- 0.09} & \num{1.0 +- 0.1} \\
Enhanced &  \num{3.8 +- 0.1 e2} & \num{1.10 +- 0.04}& \num{1.31 +- 0.06} & \num{1.5 +- 0.1} \\
Reduced & \num{6.7 +- 0.5 e2} & \num{1.14 +- 0.04}& \num{1.31 +- 0.06} & \num{0.9 +- 0.2} \\
\noalign{\vskip 1mm}  
\hline\hline
\end{tabular}
\end{table*}

We start our discussion of the experimental results by looking at the enhanced and reduced configuration. For the enhanced configuration we would expect a larger $D_\mathrm{eff}$ compared to reduced configuration because of the additive effect of the DMI and the dipolar interactions. From the measured domain widths $d$ we calculate $D_\mathrm{eff}$ and show this in~\cref{tab:out2_DMI_extract}.~\cite{Lemesh2017} Indeed, $D_\mathrm{eff}$ is larger for the enhanced configuration (\SI{1.5}{mJ.m^{-2}}) compared to the reduced configuration (\SI{0.9}{mJ.m^{-2}}). As suggested earlier, this possibly results from the modified DMI ordering in combination with the dipolar interactions that lead to \Ne{} caps.

There is another element that we have to account for in this analysis: the effect of the stacking order of the layers. For both uniform configurations, $|D_\mathrm{eff}|$ should be equal as we have simply inverted the stacking order. However, as we calculate $|D_\mathrm{eff}|$ from the measured $d$ we find that it varies by a factor of $2$ for the uniform I and II configurations as is shown in \cref{tab:out2_DMI_extract}. Moreover, there is also significant variation in the measured saturation magnetization $M_\mathrm{s}$ and anisotropy $K$ for these configurations. From this we conclude that the magnetic parameters for [Pt/Co/Ir] and [Ir/Co/Pt] vary significantly because of growth-related effects stemming from the stacking order. This seriously complicates the comparison between the different DMI configurations.

To still be able to compare the different DMI configurations, we need to extract the actual $D$ values for the enhanced and reduced configurations to conclude if there is an increase in $D_\mathrm{eff}$ for these systems. For this, we account for the growth-induced variations quantitatively by performing micromagnetic simulations in supplementary note VI. In these simulations we compare the domain wall energies in the enhanced and reduced stack with layer-dependent magnetic parameters, to a stack with the averaged parameters of both uniform stacks. From this we conclude that a hypothetical stack with these averaged magnetic parameters is a good approximation of the experimental situation, leading to errors in $D_\mathrm{eff}<0.1$~\si{mJ.m^{-2}}. To calculate $D$ for the enhanced and reduced configuration we can therefore simply average the $D_\mathrm{eff}$ of both uniform configurations, which we have also included in \cref{tab:out2_DMI_extract}.
We now compare $D_\mathrm{eff}$ of the averaged stack (\SI{1.0}{pJ.m^{-1}}) to both the enhanced and reduced stack, we find that the absolute increase of the enhanced configuration (+\SI{0.5}{pJ.m^{-1}}), and decrease of the reduced configuration (\SI{-0.1}{pJ.m^{-1}}) agree reasonably well with the values predicted in~\cref{fig:out2_over}b of $\Delta D_\mathrm{e}=$ +\SI{0.6}{pJ.m^{-1}} and $\Delta D_\mathrm{r}=$ \SI{-0.2}{pJ.m^{-1}}. Despite the growth-related complications when the building blocks of our stacks are reversed, we believe that our experimental results convincingly demonstrate that changing the sign of the DMI halfway through the stack leads to significant changes in $D_\mathrm{eff}$. Finally, in supplementary note VII we present experimental results for an $N=8$ system, with similar variations in $D_\mathrm{eff}$ as the $N=4$ system proving the wide applicability of our approach.

Summarizing the experimental part, we found changing the sign of the DMI halfway through a multilayer system is a powerful method to increase the effective DMI. With the relatively modest increase in $D_\mathrm{eff}$ of the enhanced configuration (\SI{1.5 +- 0.1}{mJ.m^{-2}}) compared to the uniform I configuration (\SI{1.3 +- 0.1}{mJ.m^{-2}}) there is still room for improvement. The largest benefit of the enhanced DMI configuration will be found in a system with a large DMI where the growth-induced variations between the opposite stacking orders are small. As there are a host of different interface combinations with a large DMI, we expect there to be many material combinations that fit this pattern.

We would now like to comment on two aspects of exploiting the \Ne{} caps to stabilize chiral spin-structures. First, the introduction of an enhanced DMI configuration does not affect the skyrmion dynamics. Although the resulting vanishing total interfacial chirality suggests that spin-orbit torques can no longer be used to drive skyrmion dynamics, this is not true for the proposed experimental stacks of [Pt/Co/Ir] and [Ir/Co/Pt].~\cite{Legrand2018,Lemesh2018,Legrand2018a,PhysRevB.98.104432} When changing the stacking order halfway through the stack, it is not only the interfacial chirality, but also the local spin-orbit torques from the individual Pt and Ir layers that are reversed.~\cite{RevModPhys.91.035004} In this case, the spin-orbit torques acting on the skyrmion are the same for both halves of the stacks, ensuring skyrmions can still be driven efficiently using an electrical current. As there are also indications that an enhanced DMI configuration can postpone the Walker-breakdown-like behaviour for both domain walls and skyrmions to much higher current densities,~\cite{PhysRevApplied.12.044031} it is therefore interesting to explore their dynamics in the case of an enhanced DMI configuration in more detail.

Second, more ideas exist that make use of dipolar interactions and magnetic parameters that vary on a layer-by-layer basis. For example, by  modifying the anisotropy, one might be able to increase the domain wall width at the top and bottom of the film to enhance the coupling with the dipolar fields and increase the skyrmion stability even further. Or, one could imagine changing the position within the stack where the DMI reverses, and thus the point at which the chirality reverses, to, for example, reduce the skyrmion Hall angle.~\cite{Legrand2018a} Perhaps it is even possible to stabilize more complex three-dimensional spin-structures such as the magnetic hopfion or skyrmion bobber by modifying individual magnetic parameters on a layer-by-layer basis.~\cite{PhysRevLett.123.147203,PhysRevLett.121.187201,Sutcliffe_2018}

In conclusion, using micromagnetic simulations we have shown that the stability of chiral spin-structures in multilayer systems can be significantly enhanced by exploiting the presence of \Ne{} caps. This can be by introducing an alternating DMI in a multilayer system, leading to increases in the effective DMI of at least \SI{0.6}{mJ.m^{-2}}. We have also shown experimental results in this direction, where we find variations in the effective DMI that agree with our predictions. These results open the way to alternative methods for the stabilization of chiral spin-structures by tailoring the magnetic interactions on a layer-by-layer basis.
\begin{acknowledgments}
This work is part of the research programme of the Foundation for Fundamental Research on Matter (FOM), which is part of the Netherlands Organisation for Scientific Research (NWO).
\end{acknowledgments}
\bibliography{references}
\end{document}


\title{Supplementary material: Stabilizing chiral spin-structures via an alternating Dzyaloshinskii–Moriya interaction}
\author{Juriaan Lucassen}
\email{j.lucassen@tue.nl}
\affiliation{Department of Applied Physics, Eindhoven University of Technology, P.O. Box 513, 5600 MB Eindhoven, the Netherlands}
\author{Mari\"{e}lle J. Meijer}
\affiliation{Department of Applied Physics, Eindhoven University of Technology, P.O. Box 513, 5600 MB Eindhoven, the Netherlands}
\author{Mark C.H. de Jong}
\affiliation{Department of Applied Physics, Eindhoven University of Technology, P.O. Box 513, 5600 MB Eindhoven, the Netherlands}
\author{Rembert A. Duine}
\affiliation{Department of Applied Physics, Eindhoven University of Technology, P.O. Box 513, 5600 MB Eindhoven, the Netherlands}
\affiliation{Institute for Theoretical Physics, Utrecht University, Princetonplein 5, 3584 CC Utrecht, the Netherlands}
\author{Henk J.M. Swagten}
\affiliation{Department of Applied Physics, Eindhoven University of Technology, P.O. Box 513, 5600 MB Eindhoven, the Netherlands}
\author{Bert Koopmans}
\affiliation{Department of Applied Physics, Eindhoven University of Technology, P.O. Box 513, 5600 MB Eindhoven, the Netherlands}
\author{Reinoud Lavrijsen}
\email{r.lavrijsen@tue.nl}
\affiliation{Department of Applied Physics, Eindhoven University of Technology, P.O. Box 513, 5600 MB Eindhoven, the Netherlands}

\date{\today}
\maketitle
\section{Methods}
\label{sec:methods}
Using \MuMax{} based micromagnetic simulations~\cite{Vansteenkiste2014} the behaviour of confined magnetic skyrmions in a circular dot (inset Fig.~1b of main paper) is investigated using a \linebreak{} [NM($2$)/FM($1$)]x$N$ system, with $N$ repeats of a \SI{2}{nm} thick ferromagnetic layer (FM) sandwiched in between \SI{1}{nm} thick non-magnetic (NM) spacer layers. For the figures in the main paper, we use the 'averaged' experimental parameters for the $N=4$ system, with a saturation magnetization $M_\mathrm{s}=1.1$~\si{MA.m^{-1}}, out-of-plane anisotropy $K=1.3$~\si{MJ.m^{-3}}, and exchange stiffness $A=10$~\si{pJ.m^{-1}}. We investigate the energy and radii of skyrmions across a range of DMI values $D$, where the DMI configuration is also varied to demonstrate the advantages of using a layer-dependent DMI in combination with \Ne{} caps. This is done using a dot with a diameter of \SI{256}{nm} and cell sizes of \SI{1}{nm} in all directions, minimizing both initial uniform and skyrmion states. In \cref{se:out28repeats} we experimentally investigate a $N=8$ system - to compare this to predictions we performed simulations for the $N=8$ system with the experimental parameters for this system which are given by $M_\mathrm{s}=1.07$~\si{MA.m^{-1}}, out-of-plane anisotropy $K=1.2$~\si{MJ.m^{-3}}, and exchange stiffness $A=10$~\si{pJ.m^{-1}}. Next, we performed additional simulations where we looked at the dependence on the number of repeats $N$, for which we used [NM($1$)/FM($1$)]x$N$ systems with $M_\mathrm{s}=1.0$~\si{MA.m^{-1}}, out-of-plane anisotropy $K=0.8$~\si{MJ.m^{-3}}, and exchange stiffness $A=10$~\si{pJ.m^{-1}}. For both the simulations for smaller dots, and those where we investigate the magnetic field dependence, we similarly used these parameters. Last, in \cref{suppl:ave} we validate the averaging approach of the main paper and for this we simulate two domain walls in a simulation box of $256 \times 32$~\si{nm^2} with periodic boundary conditions in the x and y direction of $32$ repeats. The cell sizes used are (x,y,z)=(1,8,1)~\si{nm}. We initialized two square domain walls in the x-direction of width \SI{5}{nm} with an in-plane magnetization direction exactly in between a Bloch and \Ne{} configuration.~\cite{Lucassen2019} We minimized this state and compared the energy to the uniformly magnetized state.

The experimental systems investigated in the main paper, as well as the systems presented in \cref{se:out28repeats} were DC sputter deposited using an Ar pressure of~\SI{2e-3}{mbar} on a Si substrate with a native oxide in a system with a base pressure of~\SI{4e-9}{mbar}. Pt and Ir were chosen because of their opposite interfacial DMI signs such that they favour CCW \Ne{} walls for an Pt/Co/Ir stacking ($+D$) and CW \Ne{} walls for a Ir/Co/Pt stacking ($-D$).~\cite{Ma2018,Tacchi2017,doi:10.1021/acs.nanolett.6b01593,Finizio2019,Luchaire_skyrmion} All samples were demagnetized using an oscillating decaying magnetic field applied at an $80$ to \SI{85}{\degree} angle with respect to the film normal, starting from~\SI{5}{T} decreasing in \SI{0.5}{\percent} increments with a cut-off at~\SI{0.5}{mT}. The resulting domain patterns for the individual samples were recorded using magnetic force microscopy (MFM) measurements under ambient conditions with custom-coated low-moment tips using a two-pass technique by recording the phase shift.~\cite{Hosaka1992} We extracted the periodicity using the Fourier method described in the supplementary of Ref.~\onlinecite{doi:10.1063/1.4998535}. The saturation magnetization $M_\mathrm{eff}$ and anisotropy $K$ were extracted from squid-VSM measurements using the area method.~\cite{Johnson1996}
\section{Different repeats}
\label{sec:out2_diff_repeats}
\begin{figure}
	\includegraphics[width=1\textwidth]{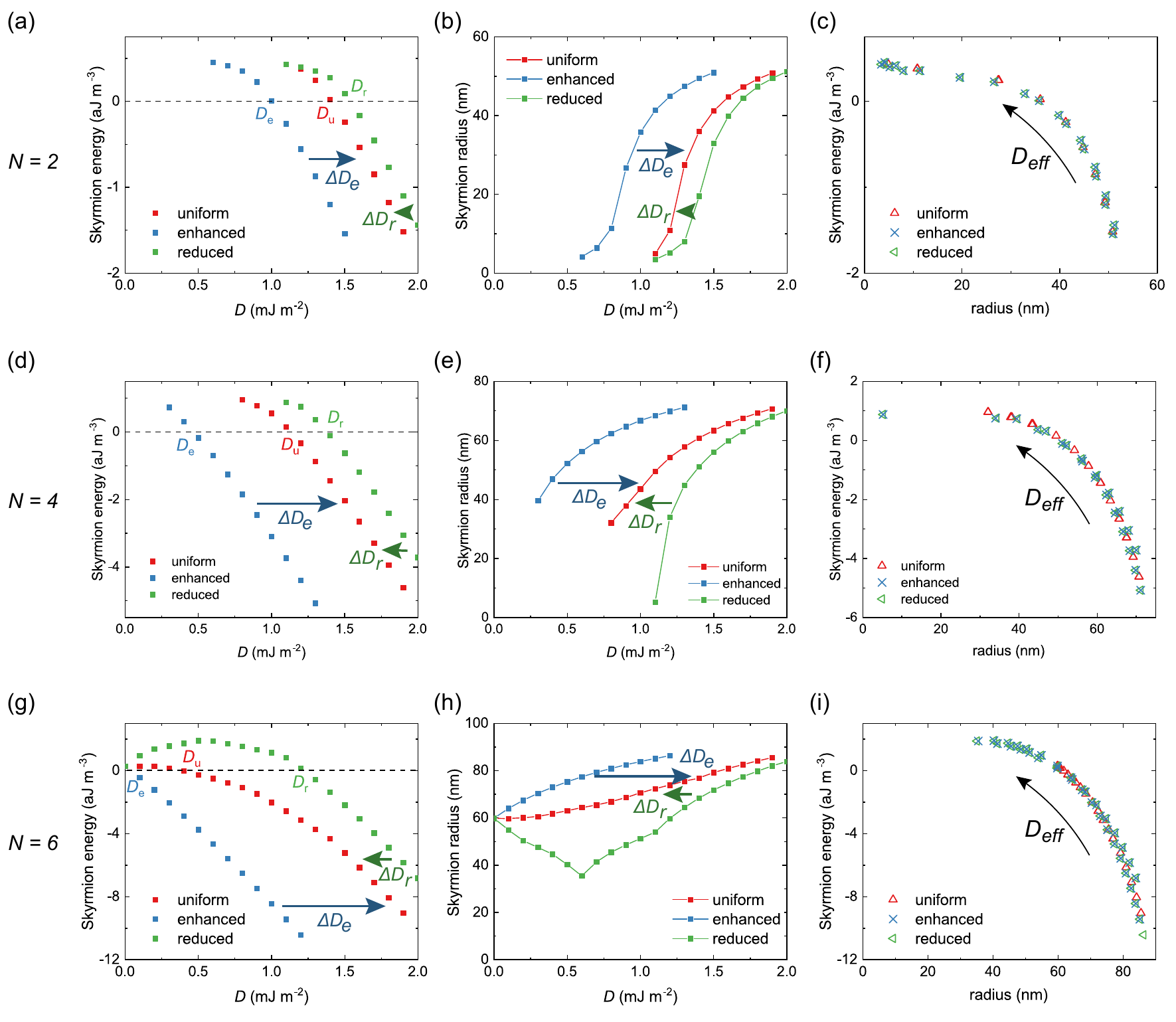}
	\caption{\label{fig:out2_allN}Behaviour of the different DMI configurations for (a-c) $N=2$, (d-f) $N=4$, and (g-i) $N=6$. (a,d,g) Skyrmion energy relative to the ground state as a function of $D$. The critical DMI values $D_\mathrm{u,e,r}$ indicate when the skyrmion energy becomes zero. (b,e,h) Skyrmion radii for three different DMI configurations determined from the position where the magnetization along the OOP axis changes sign. (c,f,i) Skyrmion energy as a function of skyrmion radius. For all figures the arrows indicate the effective DMI gain ($\Delta D_\mathrm{e}$) and loss ($\Delta D_\mathrm{r}$) as well as the direction of increasing $D_\mathrm{eff}$.}
\end{figure}

In this section we show the results of simulations where we vary the number of repeats $N$. The results for $N=2$ system are plotted in~\cref{fig:out2_allN}a-c, for the $N=4$ system in~\cref{fig:out2_allN}d-f, and the results for the $N=6$ system are shown in~\cref{fig:out2_allN}g-i. Overall, the qualitative behaviour does not change with varying $N$. From \cref{fig:out2_allN}a, d and g we find that the DMI values are reduced with increasing $N$ due to increasing influence of the dipolar interactions; for that same reason we have bigger skyrmions for larger $N$ (b, e and h). The curves for the different DMI configurations are still shifted with respect to each other, such that it is no surprise that the universal behaviour shown in Fig. 2b of the main paper is also valid for other $N$ (\cref{fig:out2_allN}c, f and i), which means that the interpretation of an effective $D$ is more widely applicable. 

\begin{figure}
	\includegraphics[width=0.6\textwidth]{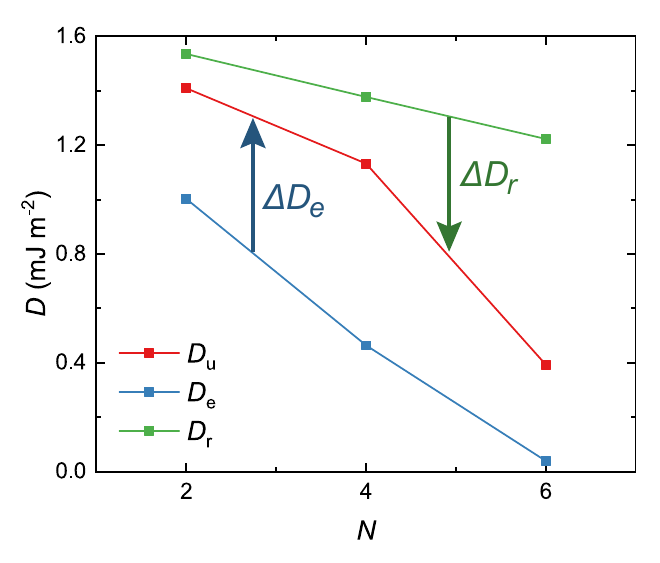}
	\caption{\label{fig:Ndep}Critical DMI values (see \cref{fig:out2_allN}) where the skyrmion energy becomes negative for the three different DMI configurations as a function of number of repeats $N$. $\Delta D_\mathrm{e}$ and $\Delta D_\mathrm{r}$ indicate the effective DMI gain and loss, respectively (see main paper for more details).}
\end{figure}

Next, we investigate the dependence of the described DMI gain ($\Delta D_\mathrm{e}$) and loss ($\Delta D_\mathrm{r}$) on the number of repeats $N$. In~\cref{fig:Ndep}, the critical DMI values $D_\mathrm{u}$, $D_\mathrm{e}$, and $D_\mathrm{r}$ at zero skyrmion energy as a function of number of repeats $N$ are given. The DMI difference between $D_\mathrm{u}$ and $D_\mathrm{e}$ increases going from $N=2$ to $4$, and then decreases again because of the complex dependence on the different contributing parameters. When starting from a uniform DMI configuration, $\Delta D_\mathrm{e}$ and $\Delta D_\mathrm{r}$ will scale with the effective dipolar fields when the initial skyrmion profile is homochiral and dominated by the DMI because the domain wall magnetization is now aligned along the dipolar fields. On the other hand, if for the uniform configuration the dipolar interactions are dominant and stabilizes two \Ne{} caps, $\Delta D_\mathrm{e}$ and $\Delta D_\mathrm{r}$ scale with $D$. Because the different regimes can be accessed by varying $N$ and $D$, the result is the complex dependence on $N$ that is demonstrated in~\cref{fig:Ndep}. 

In this interpretation, the asymmetry between $\Delta D_\mathrm{e}$ for the enhanced configuration, and $\Delta D_\mathrm{r}$ for the reduced configuration (at a given $N$) is surprising because the gain/loss in effective in-plane fields should be equal. However, there is an additional contribution from the stray fields of the domain walls themselves that prefers to align the walls in the \Ne{} cap configuration.~\cite{Tetienne2015,Hrabec2017,Bellec_2010} For the reduced DMI configuration, this provides an energy gain offsetting the large energy loss of the other two contributions.
\section{Field dependence}
\begin{figure}
	\includegraphics[width=0.6\textwidth]{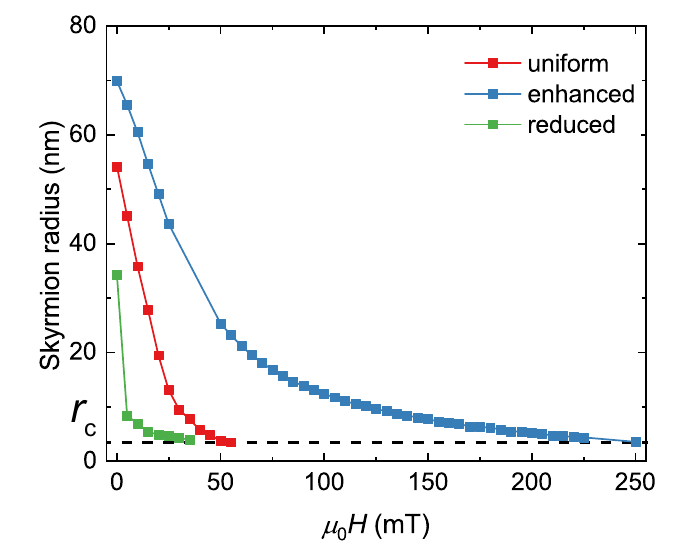}
	\caption{\label{fig:Fielddep}Radius as a function of applied magnetic field $H$ oriented anti-parallel to the skyrmion core for $D = 1.2$~\si{mJ.m^{-2}}. The dotted line indicates the smallest possible radius $r_\mathrm{c} \approx 4$~\si{nm} for a stable skyrmion.}
\end{figure}

Here we illustrate that upon introducing the different DMI configurations there is also a significant effect on the field stability of a skyrmion. The skyrmion radius is plotted as a function of a magnetic field applied anti-parallel to the skyrmion core in~\cref{fig:Fielddep}. For all configurations the radius decreases with applied magnetic field to accommodate the Zeeman energy. Yet, for the enhanced configuration skyrmions are stable up to higher fields. Additionally, we find from these simulations that the smallest possible radius $r_\mathrm{c}$ for a skyrmion does not vary noticeably with the DMI configuration, and remains constant at $\approx 4$~\si{nm} for the investigated system. It must be noted, however, that at these length scales we are close to the exchange length ($\approx 4$~\si{nm}) and cell sizes used (\SI{1}{nm}). Combined with the continuum approximation used in micromagnetic simulations, care must be therefore be taken when interpreting these results below a radius of $\sim 10$~\si{nm}.~\cite{Evans_2014,Vansteenkiste2014,PhysRevB.96.020405}
\section{Smaller diameter dot}
\label{sec:out2_smaller}
\begin{figure}[!t]
	\includegraphics[width=1\textwidth]{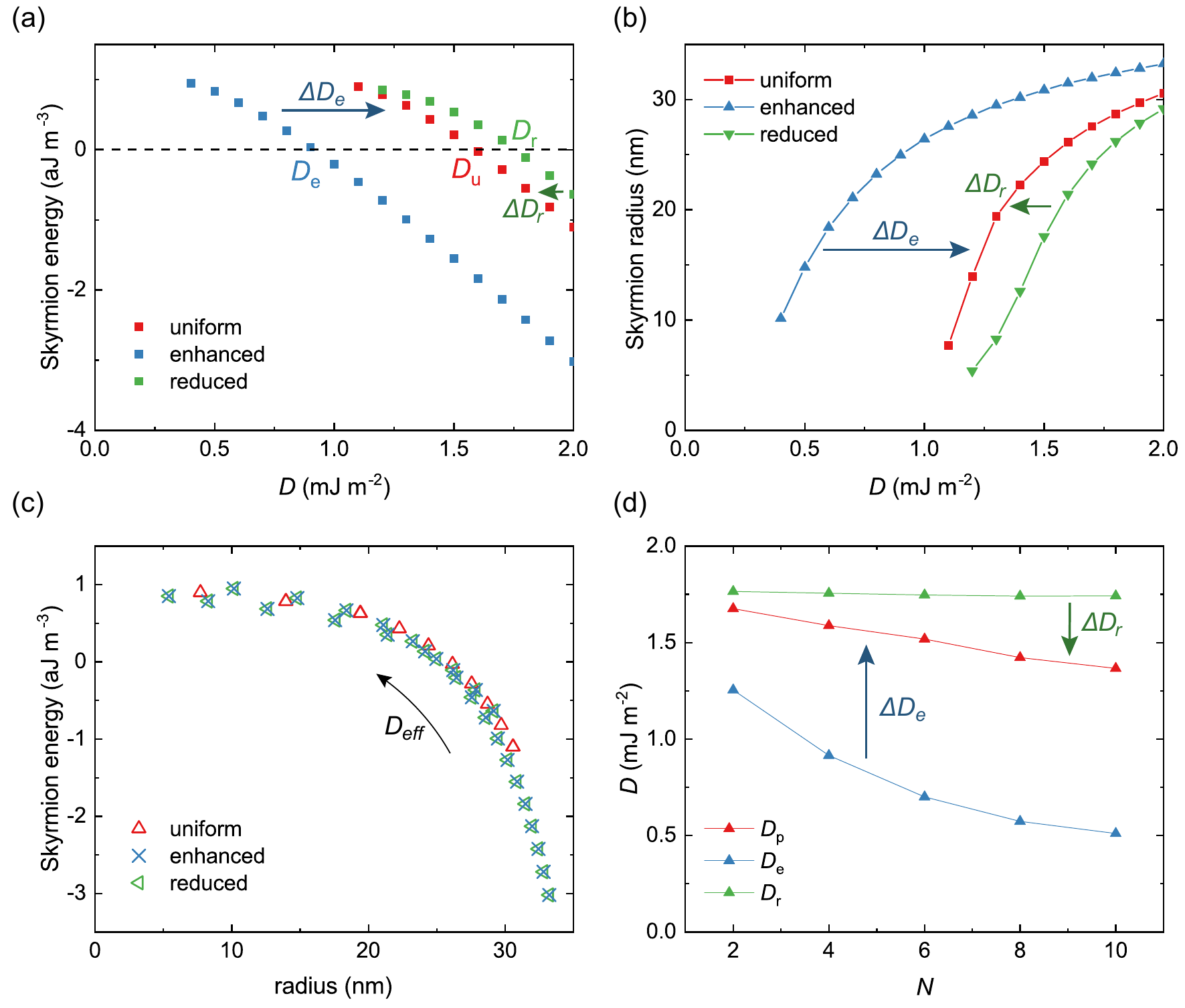}
	\caption{\label{fig:out2_smalldot}Simulations for a dot with a~\SI{128}{nm} diameter. (a-c) Results for $N=4$. (a) Skyrmion energy relative to the ground state as a function of $D$.  The critical DMI values $D_\mathrm{u,e,r}$ indicate when the skyrmion energy becomes zero. (b) Skyrmion radii for three different DMI configurations determined from the position where the magnetization along the OOP axis changes sign. (c) Skyrmion energy as a function of skyrmion radius. (d) Critical DMI values [see (a)] where the skyrmion energy becomes negative for the three different DMI configurations as a function of number of repeats $N$. For all figures the arrows indicate the effective DMI gain ($\Delta D_\mathrm{e}$) and loss ($\Delta D_\mathrm{r}$) as well as the direction of increasing $D_\mathrm{eff}$.}
\end{figure}
In order to investigate the influence of geometric confinement on the behaviour discussed in the main paper, we present results for systems with a~\SI{128}{nm} diameter dot in this section. For an $N=4$ system the skyrmion energy is plotted as a function of $D$ in~\cref{fig:out2_smalldot}a, and the skyrmion radius in~\cref{fig:out2_smalldot}b. Compared to the \SI{256}{nm} dot of \cref{fig:out2_allN}, we find that the stability region for skyrmions has shifted to larger $D$, and the skyrmion radius has decreased. These changes can be attributed to confinement effects within the dot which affect the skyrmions when their radii approach the dot radius.~\cite{PhysRevB.88.184422}

Similar to the larger dots, the physical mechanism can once again be interpreted as leading to an effective DMI through the DMI gain ($\Delta D_\mathrm{e}$) and loss ($\Delta D_\mathrm{r}$). The universal behaviour of Fig. 2b of the main paper is therefore also present for smaller dots, as we show in~\cref{fig:out2_smalldot}c. Last, in~\cref{fig:out2_smalldot}d we set out the critical DMI values as a function of number of repeats $N$, which shows the same qualitative behaviour as~\cref{fig:Ndep}. To conclude, although there is a quantitative difference due to the confinement effects within the dot, the qualitative behaviour as well as the benefits of an enhanced DMI configuration remain unaffected.
\section{Non-circular shape}
\label{sec:non_circ}
\begin{figure}
	\includegraphics[width=0.6\textwidth]{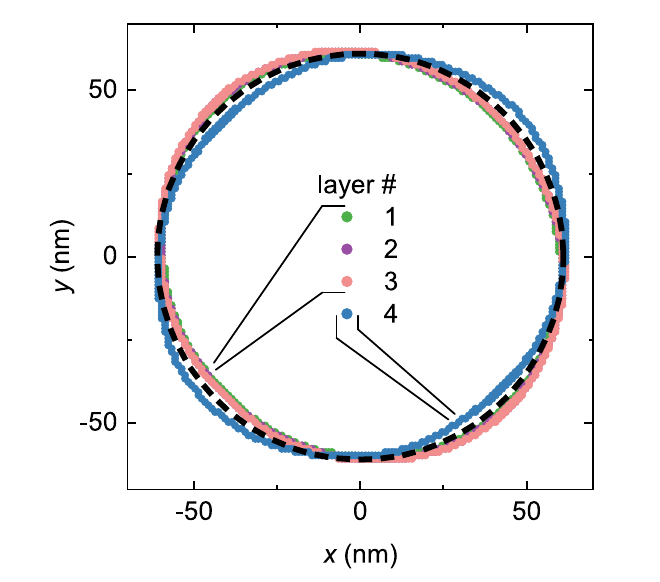}
	\caption{\label{fig:out2_noncirc}Skyrmion non-circularity (in a circular dot) for an $N=4$ system, with a uniform DMI configuration and $D=1.1$~\si{pJ.m^{-1}}. The individual layers are numbered from bottom to top; the profiles for the individiual layers are determined by locating the position where magnetization along the OOP axis changes sign, and the dashed line is the profile of a true circular skyrmion.}
\end{figure}
We note a peculiar factor of the formation of \Ne{} caps here that we believe has not been mentioned before in literature. When the effective fields from both the dipolar interaction and the DMI are approximately equal; i.e.\ when a \Ne{} cap is about to be introduced, the shape of the skyrmion is no longer circular and varies across the different repeats of the multilayer system to accommodate both the DMI and dipolar interactions. We illustrate this in~\cref{fig:out2_noncirc}, where we find that the skyrmion shape is elliptical (in a circular dot) and that the profile in the top layer has rotated its long axis with respect to the profiles in the bottom layers. Although the effect is minimal, it is worth to keep in mind when using analytical models based on axial symmetric skyrmion shapes to explore their properties.~\cite{Legrand2018a}
\section{Averaging approach validated}
\label{suppl:ave}
\begin{figure}[!t]
	\includegraphics[width=1.0\textwidth]{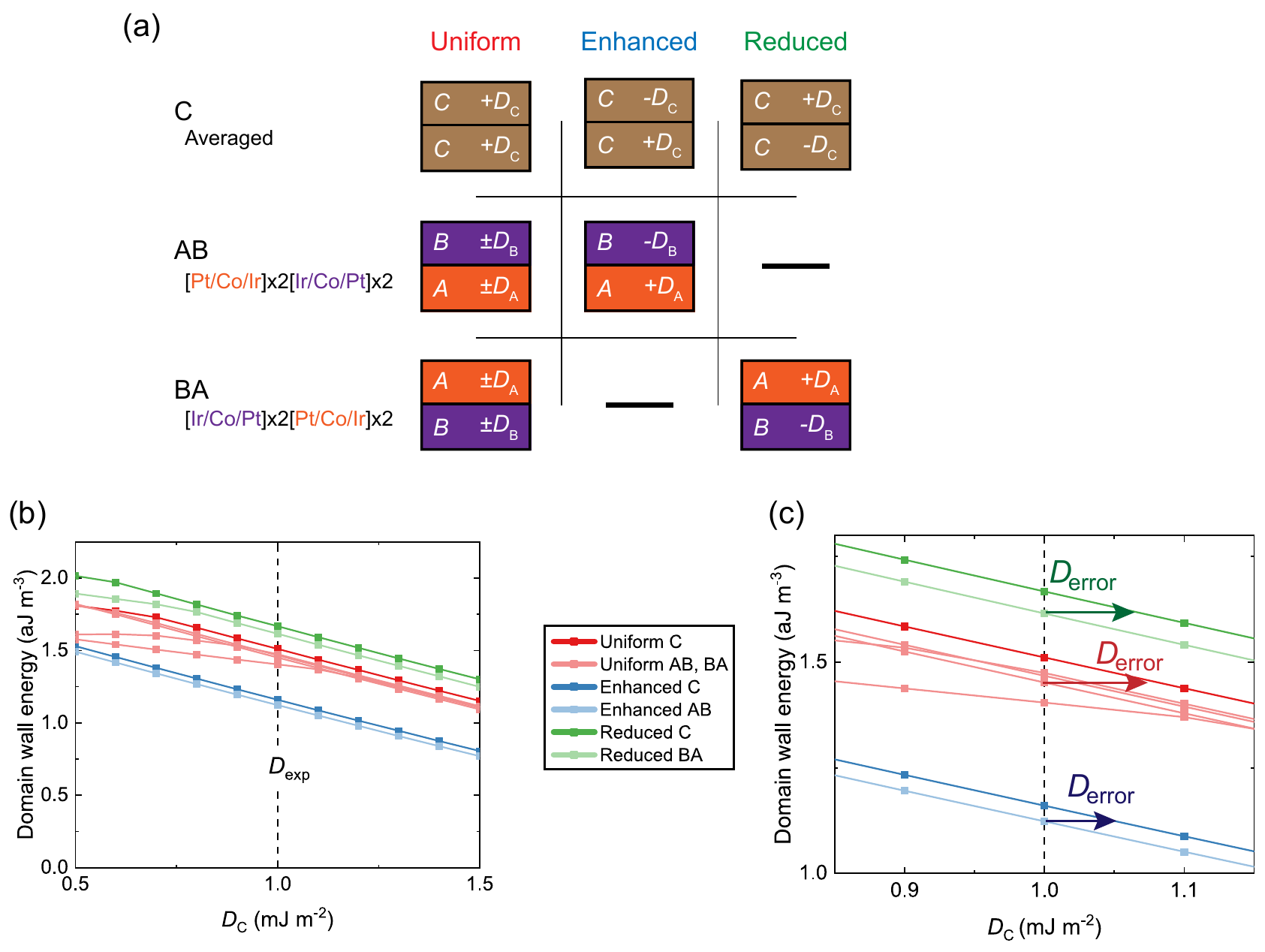}
	\caption{\label{fig:out2_averaged}Investigation of the averaging approach. (a) Table of the different simulated configurations. We have the average configurations (C), which use 'average parameters' of the system (see main paper), and the experimental stacks with a thickness dependent $M_\mathrm{s}$ and $K$, with AB stacking the enhanced configuration and BA stacking the reduced configuration. For the exact parameter values used in the simulations, see \cref{tab:out2_averaged}. (b) Domain wall energy as a function of average DMI value $D_\mathrm{C}$ for the configurations shown in (a). $D_\mathrm{A}-D_\mathrm{B}$ is always kept to \SI{0.7}{mJ.m^{-2}}, with $D_\mathrm{C}=(D_\mathrm{A}+D_\mathrm{B})/2$. $D_\mathrm{exp}$ is the experimental 'averaged' value, and corresponds with the C value in \cref{tab:out2_averaged}. (c) Zoomed version of (b), where we indicate the $D$ shifts between the averaged and full stacks with $D_\mathrm{error}$.}
\end{figure}
\begin{table}
\caption{\label{tab:out2_averaged}Micromagnetic parameters used for simulations displayed in \cref{fig:out2_averaged}. A and B are the experimental values for Pt/Co/Ir and Ir/Co/Pt, respectively (see Table 1 of main paper). C corresponds to the average of Pt/Co/Ir and Ir/Co/Pt.}
\begin{tabular}{l cccc }
\hline\hline
\noalign{\vskip 1mm}  
         &   Stack & $M_\mathrm{s}$ (\si{MA.m^{-1}}) & $K$ (\si{MJ.m^{-3}}) & $D$ (\si{mJ.m^{-2}}) \\   
     \noalign{\vskip 1mm} 
A & [Pt/Co/Ir]x$2$ & \num{0.9}& \num{1.02} & \num{1.3} \\
B &  [It/Co/Pt]x$2$ & \num{1.3}& \num{1.6} & \num{0.6} \\
C & Average of A and B & \num{1.1}& \num{1.31} & \num{1.0}\\

\noalign{\vskip 1mm}  
\hline\hline
\end{tabular}
\end{table}
In this section we verify, using micromagnetic simulations, that the averaging approach advocated in main paper only leads to errors in the extracted $D_\mathrm{eff}$ of at maximum, \SI{0.1}{mJ.m^{-2}}, well within the experimental error bars. To demonstrate this, we simulate several configurations as indicated in \cref{fig:out2_averaged}a, using the values given in \cref{tab:out2_averaged}. There are three different stack configurations: i) the 'averaged stack' C which uses the averaged micromagnetic parameters of Pt/Co/Ir and Ir/Co/Pt (see Table 1 of main paper), ii) the enhanced stack which we call AB because it contains $2$ repeats of Pt/Co/Ir (A) and then $2$ repeats of Ir/Co/Pt (B), and iii) the reduced stack BA with the stacking order reversed. For each of these configurations, we vary the sign of the DMI to simulate the uniform, enhanced and reduced DMI stacking order to investigate the effect of the averaging approach using micromagnetic simulations. Rather than investigating the skyrmion energy, we now look at the domain wall energy because skyrmions are not stable at the experimental DMI values for all DMI configurations (details on the simulations can be found in \cref{sec:methods}). 

When looking at results for these different configurations, we are trying to verify that the domain wall energies for the averaged simulations (C) do not vary significantly from the energies for the exact experimental configurations (AB and BA). If they are approximately equal, the averaging approach is a good approximation of the experimental situation with layer-dependent magnetic parameters. Specifically, we are trying to verify that the uniform AB/BA situation yields results that are approximately equal to the uniform C configuration. If this is true, calculating an average DMI as we do in the main manuscript is allowed. Comparing this averaged DMI to the experimental $D_\mathrm{eff}$ then involves ensuring the enhanced C and enhanced AB configuration give approximately equal domain wall energies and the reduced C and reduced BA situation similarly give approximately equal energies. If this is the case, we have verified for the experimental stacks that the averaging approach gives similar results to a situation with layer-dependent magnetic parameters.  

We plot the energy for all these configurations in \cref{fig:out2_averaged}b as a function of the average DMI $D_\mathrm{c}$. The darker lines correspond to the 'averaged' configuration C as well as the simulations presented in the main paper. The lighter lines correspond to the full simulated stacks, including a thickness dependent $M_\mathrm{s}$ and $K$. As can be seen, the results of configuration C (averaged) line up very well with the exact experimental configurations (AB and BA), except for the uniform stack at low $D_\mathrm{c}$ values. Here, variations in the coupling of the stray fields to the domain walls are significantly different because of the layer dependent parameters. The fact that the curves for configuration C (averaged), AB, and BA are almost identical for $D_\mathrm{c}>0.9$~\si{mJ.m^{-2}} suggests that the averaging approach is valid in this region; if this were not the case, there would be significant variations in the domain wall energy and configurations C, BA and AB would describe a different physical system.

For the experimental $D_\mathrm{c}=D_\mathrm{exp}=1.0$~\si{mJ.m^{-2}} we find that the curves overlap very well, and we can quantify the error of the 'averaging' approach using the zoom-in shown in \cref{fig:out2_averaged}c. Here, we indicate these errors in the extracted $D_\mathrm{eff}$ by $D_\mathrm{error}$, based on the difference in $D_\mathrm{c}$ values between the full experimental stack (AB and BA) and the 'averaging approach' (C). We find in all cases that $D_\mathrm{error}<0.1$~\si{mJ.m^{-2}}, well within the experimental error bars. This means that the 'averaging' approach described in the main paper is fully justified.
\section{8 repeats}
\label{se:out28repeats}
\begin{figure}
	\includegraphics[width=1\textwidth]{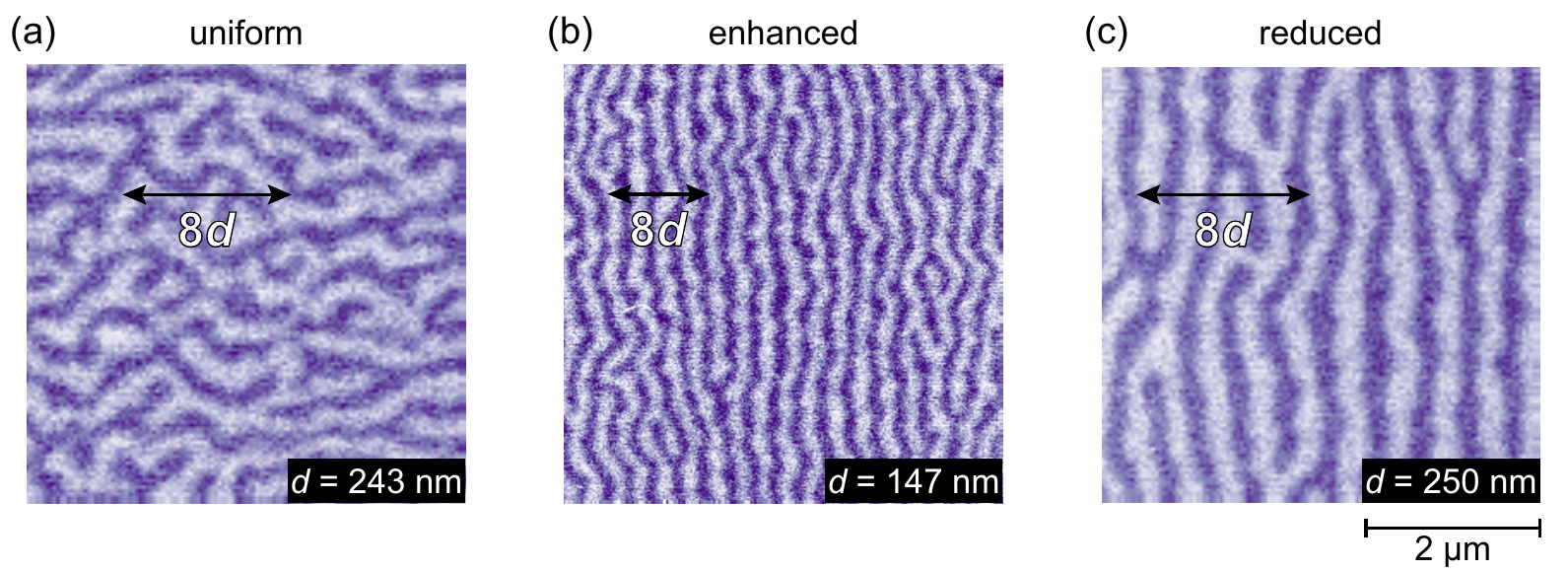}
	\caption{\label{fig:out2_MFM2}MFM images (for $N=8$ systems) of a demagnetized domain state for (a) the uniform stack, (b) the enhanced stack, (c) and the reduced stack. The scale bar at the bottom right holds for all figures. The arrows indicate the average domain width $d$, which we also show in the bottom right of each scan.
}
\end{figure}
\begin{table}
\caption{\label{tab:out2_DMI_extract2}%
Analysis of the domain widths $d$ for the three different stacks extracted from~\cref{fig:out2_MFM2}. From these values the effective DMI $|D_\mathrm{eff}|$ was calculated using the extended stripe domain model with $A=10$~\si{pJ.m^{-1}}.~\cite{Lemesh2017}}
\begin{tabular}{l cccc }
\hline\hline
\noalign{\vskip 1mm}  
         &   $d$ (\si{nm}) & $M_\mathrm{s}$ (\si{MA.m^{-1}}) & $K$ (\si{MJ.m^{-3}}) & $|D_\mathrm{eff}|$ (\si{mJ.m^{-2}}) \\   
     \noalign{\vskip 1mm} 
Uniform & \num{243 +- 5} & \num{0.84 +- 0.03}& \num{0.86 +- 0.03} & \num{1.44 +- 0.09} \\
Enhanced &  \num{147 +- 7} & \num{1.04 +- 0.04}& \num{1.14 +- 0.05} & \num{1.5 +- 0.1} \\
Reduced  & \num{2.5 +- 0.1e2} & \num{1.10 +- 0.04}& \num{1.27 +- 0.06} & \num{0.8 +- 0.2}\\

\noalign{\vskip 1mm}  
\hline\hline
\end{tabular}
\end{table}
In this section we demonstrate that the experimental increase and decrease in $D_\mathrm{eff}$ observed in the main paper for $N=4$ systems can also be found for $N=8$ systems. To check this, we once again fabricated Ta($4$)/Pt($2$)/X/Ta($4$) stacks, but with X for each stack now given by: \vspace{2.5mm}\newline{}
\textbf{Uniform:} $\overbrace{\mathrm{[Pt}(1)\mathrm{/Co}(1)\mathrm{/Ir}(1)\mathrm{]}}^{+D}$x$8$, \vspace{2.5mm}\newline{}
\textbf{Enhanced:} $\overbrace{\mathrm{[Pt}(1)\mathrm{/Co}(1)\mathrm{/Ir}(1)\mathrm{]}}^{+D}$x$4$$\overbrace{\mathrm{[Ir}(1)\mathrm{/Co}(1)\mathrm{/Pt}(1)\mathrm{]}}^{-D}$x$4$, \vspace{2.5mm}\newline{}
  \textbf{Reduced:} $\overbrace{\mathrm{[Ir}(1)\mathrm{/Co}(1)\mathrm{/Pt}(1)\mathrm{]}}^{-D}$x$4$$\overbrace{\mathrm{[Pt}(1)\mathrm{/Co}(1)\mathrm{/Ir}(1)\mathrm{]}}^{+D}$x$4$.\newline{}\vspace{2.5mm}\newline{}
In~\cref{fig:out2_MFM2} we present the MFM scans, and in~\cref{tab:out2_DMI_extract2} the quantitative analysis following the same procedure as presented in the main paper. Similar to the $N=4$ stack we find a large difference in the effective DMI between the enhanced and reduced stack with similar values of $M_\mathrm{s}$ and $K_\mathrm{eff}$, completely in line with the predictions and the results of the $N=4$ system. Furthermore, simulations suggest that variations in effective DMI for the $N=8$ system should be slightly larger than the $N=4$ system (not shown). The difference in effective DMI for the $N=8$ system between the reduced and enhanced configuration is predicted to be approximately \SI{1.0}{mJ.m^{-2}} compared to \SI{0.8}{mJ.m^{-2}} for the $N=4$ system. Experimentally, the variation for the $N=4$ system of about \SI{0.6}{mJ.m^{-2}} agrees with the prediction, but the variation for $N=8$ of \SI{0.7}{mJ.m^{-2}} is on the low side when compared to the simulations. Again, however, we find that the uniform and enhanced configuration have an approximately equal $D_\mathrm{eff}$ that is the same as that of the $N=4$ stack, thus suggesting that growth-induced changes play a major role here as well. It is therefore difficult to exactly quantify the $D_\mathrm{eff}$ variations when growth variations along the stack thickness also play a role here.
\bibliography{../references}